\newcommand{\beq}{\begin{equation}}

\newcommand{\eeq}{\end{equation}}
\newcommand{\bea}{\begin{eqnarray}}
\newcommand{\eea}{\end{eqnarray}}
\newcommand{\bear}{\begin{eqnarray*}}
\newcommand{\eear}{\end{eqnarray*}}
\newcommand{\lb}{\label}
\newcommand{\rf}[1]{(\ref{#1})}

\def\bra#1{\langle #1|}
\def\ket#1{|#1\rangle}

\documentclass[12pt]{iopart}
\usepackage{epsfig} 

\usepackage{psfrag}
\usepackage{graphicx}
 
\begin{document}

\title
{Exactly solvable interacting vertex models}

\author{Francisco C. Alcaraz$^1$ and Matheus J. Lazo$^2$} 
\address{$^1$Instituto de F\'{\i}sica de S\~ao Carlos, 
Universidade de S\~ao Paulo, Caixa Postal 369, 13560-590, 
 S\~ao Carlos, SP,
Brazil}
\ead{alcaraz@if.sc.usp.br}
\address{$^2$Centro Tecnol\'ogico de Alegrete, Universidade Federal de Santa
  Maria/Unipampa,\\ Alegrete, RS, Brazil\\ Departamento de F\'\i sica, Universidade Federal de Santa
  Maria, 97111-900, \\ Santa Maria, RS, Brazil}

\ead{lazo@smail.ufsm.br}

\begin{abstract}

We introduce and solve 
a special family of integrable interacting vertex models 
that generalizes the well known six-vertex model. 
 In addition to the usual nearest-neighbor interactions among the vertices, there 
exist extra hard-core interactions among pair of vertices at larger distances. The associated 
row-to-row transfer matrices  are diagonalized by using the recently introduced 
matrix product {\it ansatz}. Similarly as the relation of the six-vertex model 
with the XXZ quantum chain, the row-to-row transfer matrices of these knew 
models are also the generating functions of an infinite set of commuting 
conserved charges. Among these charges we identify the integrable generalization 
of the XXZ chain that contains hard-core exclusion interactions among the spins. 
These quantum chains already appeared in the literature. The present paper explains 
their  integrability.

\end{abstract}

\noindent{\it Keywords\/: Integrable spin chains (vertex models), Solvable lattice models, Quantum integrability (Bethe Ansatz) }

\date{\today}
\maketitle

\section{Introduction}\label{section1}

The six-vertex model was introduced by Pauling in order to explain the
residual entropy of the ice at zero temperature. The model turns out to be 
of great interest for the physics and mathematics of many body interacting 
systems, due to its 
 integrability.  The row-to-row transfer matrix of the six-vertex 
model is the generating function for an infinite set of commuting non-trivial 
charges in involution \cite{tarasov}. The XXZ quantum chain being one of these charges. 
For a quantum 
system to be integrable  its Hamiltonian should belong to an infinite set 
of commuting operators. The  integrability of the  XXZ quantum chain is then 
a consequence of the infinite number of commuting charges generated by the 
six-vertex model. For this reason the  
six-vertex model is considered as a paradigm of  integrability in 
statistical mechanics \cite{lieb}-\cite{gaudin}

The connection of vertex models and quantum chains was also observed in a great 
variety of general vertex models and quantum chains thanks to the development 
of the quantum inverse scattering method [QISM] \cite{fadeev}-\cite{sklyanin}. 
The QISM allows 
the construction of the vertex models and the associated  quantum chains from 
the solutions of the Yang-Baxter equations (see \cite{korepin}-\cite{schlot1} for reviews). Basically we 
should expect that associated to any integrable quantum chain in one 
dimension should exists a vertex model whose row-to-row transfer matrix is 
the generating function for the quantum chain. 

On the other hand, along the recent years, it has been shown that it is 
possible to generalize several known  integrable quantum chains by 
preserving their  integrability. These generalizations are obtained 
through the introduction of suitable hard-core exclusion interactions. 
Examples of these 
generalizations were obtained for the XXZ quantum chain \cite{alcbar1,maquire}, 
spin-1 Fateev-Zamolodchikov and Izergin-Korepin models \cite{alcbar3}, 
$SU(N)$ Sutherland and Perk-Schultz models \cite{alcbar1,alcbar5} and the 
Bariev model \cite{schlot2,schlot3}. Some of these extended quantum chains appear in the 
description of the stochastic dynamics of in the asymmetric exclusion problem 
with particles with extended sizes \cite{sazamoto,alcbar5},\cite{alcbar6}-\cite{schonherr}.
The exact solution of these quantum chains were not obtained by the QISM 
but directly by the coordinate Bethe {\it ansatz} \cite{bethe} or by the 
matrix product ansatz \cite{alclazo1}-\cite{alclazo4}. Consequently, the two dimensional vertex 
model generating these quantum chains are not known, and we cannot explain 
their  integrability, up to now. We expect that similar to the 
quantum chains the associated vertex models should have, beyond the 
nearest neigbour interactions imposed by the lattice connectivity, also  additional 
hard-core interactions.

Recently an  interacting five-vertex model in a square lattice was 
introduced and solved \cite{ice}. In this model there exist  
hard-core interactions along  the diagonal of the square lattice. The exact 
solution was obtained only for the eigenspectra of the diagonal-to-diagonal 
transfer matrix, that distinctly from the row-to-row transfer matrix does 
not generate an infinite set of conserved charges.

In this paper we are going to introduce a family of  integrable 
interacting vertex models that generalize the six-vertex model and explains 
 the  integrability of the XXZ quantum chains with hard-core interactions. 
The row-to-row transfer matrices associated to these vertex models generate 
an infinite set of conserved charges, that include the Hamiltonians of the  
hard-core interacting quantum chains. 

In the six-vertex model the interactions are between nearest neighbour 
vertices and are ruled by the geometrical connectivity of the allowed 
vertex configurations of the lattice. The interacting energy among two 
vertices is zero for allowed configurations or infinite otherwise. The 
family of interacting vertex models we introduce in this paper have 
additional interactions. Besides the usual nearest-neighbour interactions, 
imposed by the lattice connectivity, there exist also interactions 
among vertices at larger distances.

The exact solution of transfer matrices associated to vertex models or quantum
Hamiltonians are usually obtained through the Bethe ansatz
\cite{bethe} in its several formulations. The ansatz asserts
that the amplitudes of the eigenfunctions of these operators are given by a
sum of appropriate plane waves. Instead of making use of the Bethe ansatz, the
solution we derive will be obtained through a matrix product ansatz
(MPA). 

Under the general name of MPA several methods were introduced in the literature 
along the years. The first formulation was done for the description of the 
ground-state eigenfunction of some special nonintegrable quantum chains, the 
so called valence-bond solid models \cite{affleck}-\cite{kluemper1}. 
MPA becomes also a successfull tool for the exact calculation of the 
stationary probability distribution of some stochastic one dimensional 
systems \cite{derrida1}-\cite{dasmahapatra}. An extension of this last 
MPA, called dynamical MPA was introduced in \cite{schutz1,schutz2} and 
extended in \cite{popkov1}. This last ansatz gives the time-dependent 
probability distribution for some  integrable systems. 

The MPA we are going 
to use in this paper, in order to solve the new family of integrable 
vertex models, was introduced in \cite{alclazo1}-\cite{alclazo4}. 
This ansatz was applied with success in the evaluation of the eigenspectra of 
several  integrable quantum Hamiltonians \cite{alclazo1}-\cite{alclazo3}, 
transfer matrices \cite{ice,lazo} and the time-evolution operator of stochastic 
systems \cite{alclazo4}.
According to this ansatz, the
amplitudes of the eigenfunctions are given in terms of a product of matrices
where the matrices obey appropriated algebraic relations. In the case of
the Bethe ansatz the spectral parameters and the amplitudes of the
plane waves are fixed, apart from a  normalization constant, by the eigenvalue
equation of the Hamiltonian or transfer matrix. On the other hand, in the MPA 
the eigenvalue
equation fixes the commutation relations of the matrices defining the ansatz. 

The layout of the paper is as follows. In section $2$ we introduce the
interacting vertex models we are going to solve.  In section $3$
we present the row-to-row transfer matrices associated to the  vertex models. 
In section 4 we explain the MPA and give the exact solution of the vertex 
models. In section 5 we derive the associated quantum chains that commute with 
the row-to-row transfer matrices, solved in section 4. Finally in section 6 
we conclude the paper with a general discussion.


\section{The interacting four-vertex model} \label{section2}

\begin{figure}[ht]
\centerline{
\begin{picture}(270,130)
\put(0,0){\epsfxsize=300pt\epsfbox{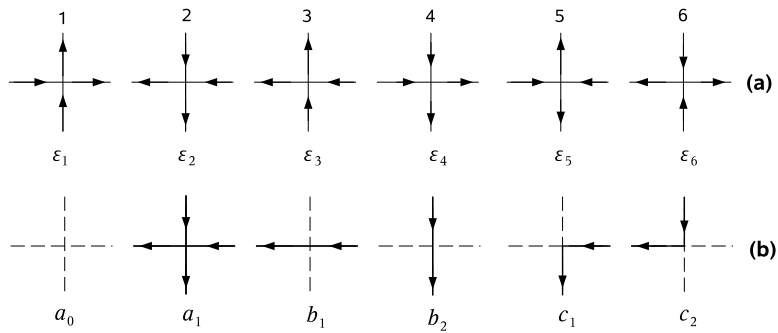}}
\end{picture}}
\caption{The six possible vertex configurations allowed by the ice rules, with
  their respective energies and fugacities. In (a) we draw all arrows and in (b) 
  we give a representation where only the arrows pointing down and left are drawn.}
\label{fig1}
\end{figure}

The family of interacting vertex models we introduce are defined on a square
lattice with M rows and L columns and toroidal boundary conditions. At each
horizontal (vertical) lattice bond we attach an arrow pointing to the left or
right (up or down) direction. These arrows configurations can be equivalently
described by the vertex configurations of the lattice. A vertex configuration
at a given site (center) is formed by the four arrows attached to its
links. Similar to the asymmetric six-vertex model \cite{Nolden,BukmanShore} we
impose that the allowed arrow configurations only contain vertices satisfying
the ice rules: two of the arrows pointing inward and the other two pointing
outward of its center. For the standard six-vertex model there are six
possible configurations for the vertices. These configurations are showed in
Fig.~\ref{fig1}a with their respective energies $\varepsilon_1, \ldots,
 \varepsilon_6$. The partition
function is given by the sum of all possible vertex configuration with the
Boltzmann weights given by the product of the fugacities $e^{-\beta
  \varepsilon_i}$ ($i=1,\ldots,6$) of the vertices. In Fig.~\ref{fig1}b, a more convenient
notation is introduced, in which we only drawn, with the corresponding fugacities 
$a_0$, $a_1$, $b_1$, $b_2$, $c_1$ and $c_2$,  the arrows pointing to the left
or down of the center defining the vertex. 


 As a consequence of the ice rules and the periodicity 
of the lattice the  number of 
down arrows $n$ (or up arrows $L-n$) is conserved on each  row of vertical
bonds.

It is important to notice that the ice rules imply the existence of
interactions among the nearest-neighbor vertices. These interactions have a
zero or infinite value. For example (see Fig.~\ref{fig1}b), the vertex with fugacity
$a_0$ has an infinite interaction energy if the vertex on its right side is
one of the vertices ($a_1, b_1, c_2$) or the vertex on its left side is one of the
vertices ($a_1, b_1, c_1$). The model with no extra interactions, besides
those given by the connectivity of the vertices on the lattice, is the well
known six-vertex model. The partition function of the model is given by the
sum of all possible vertex configurations with the Boltzmann weights given by
the product of the fugacities of the vertices. The six-vertex model is exactly
integrable for arbitrary values of the fugacities, and is considered a
prototype of an exact solvable model 
\cite{lieb}-\cite{baxter},\cite{Nolden,BukmanShore}.  

We consider a special family of interacting four-vertex models. Besides the
previously mentioned nearest-neighbor interactions (infinite or zero) imposed
by the lattice connectivity, the models also contain interactions among pairs
of vertices at larger distances. The allowed four vertex configurations, with their
respective configurations are the vertices $1$, $3$, $5$ and $6$ shown in
Fig.~\ref{fig1}. Contrary to the six-vertex model the vertex configurations with
fugacities $a_1$ and $b_2$ are forbidden. Such interacting
four-vertex models are labeled by two fixed positive integer $s_1$ and $s_2$
that may take the values $s_1,s_2=1,2,\ldots$. These parameters specify the
additional interactions among the vertices that occur when there are vertices
of types $c_1$ and $c_2$ at distances equal to $s_1$ or $s_2$, in lattice
units along the horizontal lines of the square lattice. A pair of
vertices at distance ($l=1,2,...$), in units of lattice spacing, along this
horizontal line interacts following the rules\footnote{For simplicity we denote a given 
vertex by its fugacity, and a pair of vertices where the vertex fugacity $c_1$ is on 
the left (right) of $c_2$ by $c_1 - c_2$ ($c_2 - c_1$), respectively.}

\begin{figure}[ht]
\centerline{
\begin{picture}(270,130)
\put(0,0){\epsfxsize=330pt\epsfbox{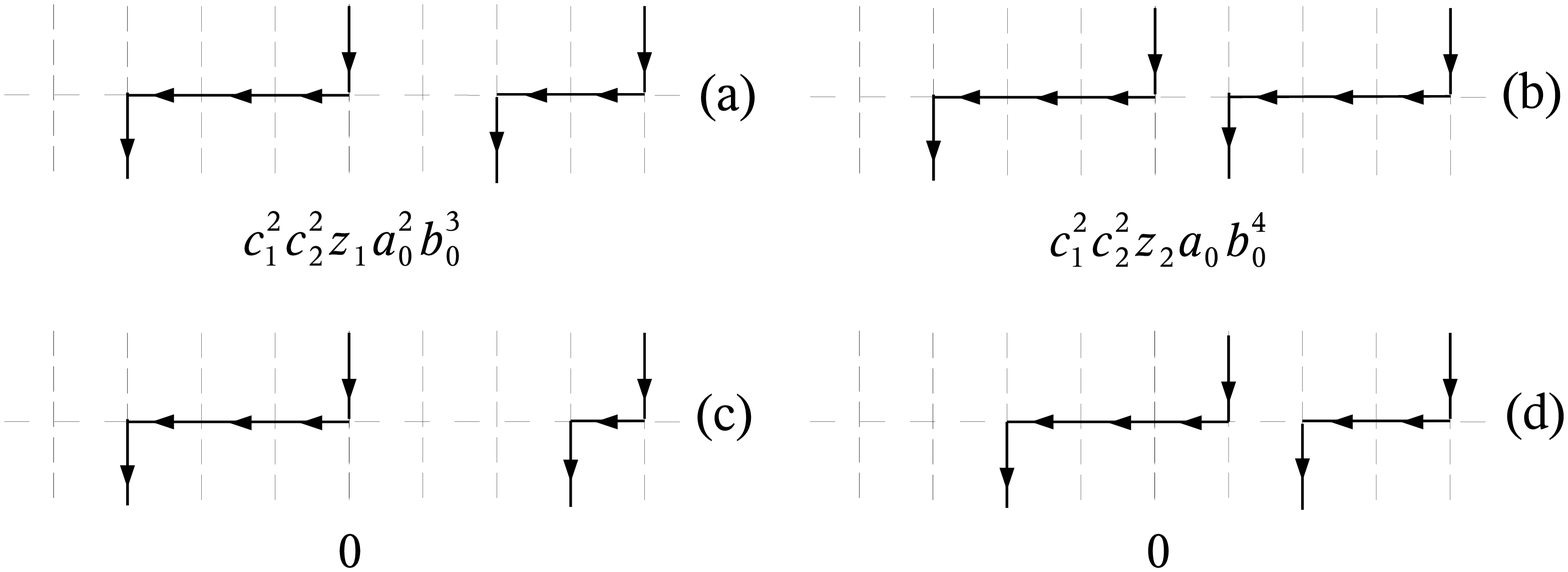}}
\end{picture}}
\caption{
Configurations containing two arrows in a lattice of size $L=9$, for the 
model with $s_1=2$ and $s_2=1$. The configurations $(a)$ and $(b)$ are allowed and their respective Boltzmann weights contributions are shown. Configurations 
$(c)$ is not allowed due to the occurence of a pair $c_1-c_2$ at distance $1<s_1$. configuration $(d)$ is also not allowed because the two vertices $c_2$ are at distance $3\leq s_1+s_2$.}
\label{fig2}
\end{figure}

\begin{itemize}
\item[(a)] The interaction energy is zero if one of the vertices is $a_0$ or $b_1$.

\item[(b)] If the pair of vertices is of type $c_1 -c_1$ or $c_2-c_2$ the 
interaction energy is infinity if $l\leq s_1+s_2$.

\item[(c)] If the pair of vertices is of type $c_1-c_2$ ($c_2-c_1$) the interaction 
energy is infinity if $l <s_1$ ($l<s_2$) and zero if $l>s_1$ ($l>s_2$). In the 
special case where $l=s_1$ ($l=s_2$) the interaction energy is finite and 
produces, besides the fugacity of the vertices, a contribution to the 
Boltzmann weight $z_1$ ($z_2$) given by 

\end{itemize}
\beq
\label{e1}
z_1=\frac{b_1b_2}{c_1c_2} \;\;\;\;\; \mbox{and} \;\;\;\;\; z_2=\frac{a_0a_1}{c_1c_2}.
\eeq
 The use of the missing fugacities $a_1$ and $b_1$ to denote the interactions
$z_1$ and $z_2$ in \rf{e1} is convenient for the forthcoming analysis.
In Fig.~\ref{fig2} we show some examples of allowed and not allowed configurations 
containing two arrows in the model with 
$s_1 =2$ ans $s_2=1$. 

 In general,
the contribution of a given pair of vertices is zero if the pair is not
allowed (infinite interaction energy) or is given by the product of their
fugacities (zero interaction energy). The exception to this rule happens when
we have the pair $c_1-c_2$ ($c_2-c_1$) at the distance $s_1$
($s_2$) along the horizontal line (see Fig.~\ref{fig2}). In this case, from \rf{e1},
the total contribution to the Boltzmann weight is given by 
$c_1c_2z_1 = b_1b_2$ ($c_1c_2z_2 = a_0a_1$).  

We can also extend our model to the cases where the interaction parameters 
$s_1,s_2$ may take the value zero. In these cases the interaction follows the previous rules 
(a)-(c) but instead of \rf{e1} we have the generalized form
\beq \label{e2}
z_1=\frac{b_1b_2}{c_1c_2}b_1^{-\delta_{s_1,0}} 
 \;\;\;\;\; \mbox{and} \;\;\;\;\; z_2=\frac{a_0a_1}{c_1c_2}a_0^{-\delta_{s_2,0}},
\eeq
where $\delta_{s,0}$ is the usual Kronecker delta. In this generalization we may have the 
pair of vertices $c_1-c_2$ ($c_2-c_1$) at distance $s_1=0$ ($s_2=0$). On this case the total 
contribution comming from the pair is now $c_1c_2z_1 = b_2$ ($c2c_1z_2=a_1$). 
The models where $s_1=0$ and $s_2\neq 0$ ($s_2=0$ and $s_1\neq0$) are equivalent 
to interacting five vertex models where only the vertex $a_1$ ($b_2$) of 
Fig.~\ref{fig1} is not allowed. 
The interactions on these models, according to  rule (a), forbid  arrows at
 distances 
smaller than $s_2+1$ ($s_1+1$).   Moreover the pair of vertices $c_1-c_2$ 
($c_2-c_1$) interacts as in rule (c). 
The special case where $s_1=s_2=0$ recovers the standard six-vertex model, with 
interactions given only by the lattice connectivity and the ice rules.

In summary, the general vertex models defined by the rules (a)-(c) and by \rf{e2}, contains
interacting four-vertex models ($s_1\neq0, s_2\neq 0$), interacting five-vertex models 
($s_1=0,s_2\neq0$ or $s_1\neq 0,s_2=0$) and the standard six-vertex model 
($s_1=s_2=0$).

\begin{figure}[ht]
\centerline{
\begin{picture}(250,150)
\put(-30,0){\epsfxsize=350pt\epsfbox{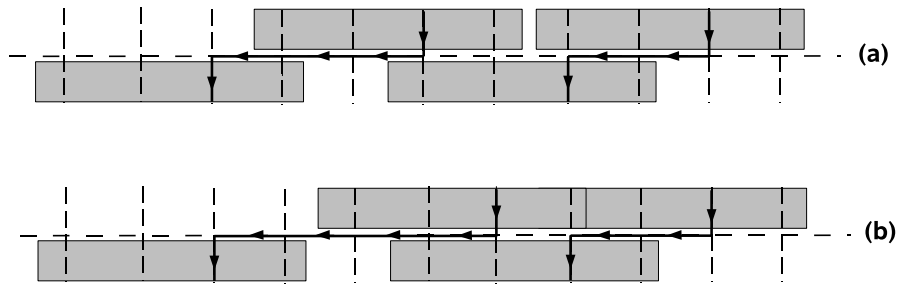}}
\end{picture}}
\caption{Examples of an allowed (a) and not allowed (b) configuration of the
  interacting four-vertex model with $s_1=2$ and $s_2=1$. The vertical arrows
  have an effective hard-core size $s=s_1+s_2+1=4$.}
\label{fig3}
\end{figure}

It is important to notice  that the interactions, due to rule (b), forbid 
two 
vertical arrows  at distances smaller than $s_1+s_2+1$
($s_1,s_2=0,1,2,\ldots$). This can  be interpreted as if the vertical arrows have an
effective size $s_1+s_2+1$ ($s_1,s_2=0,1,2,\ldots$), in units of lattice
spacing. A vertical arrow on a given link has hard-core interactions that
exclude the occupation of other vertical arrows at the link itself as well as the $s_1$
 nearest links on its left and $s_2$ nearest links on its right. 
In Fig.~\ref{fig3}  we
represent pictorically an allowed (a) and a not allowed (b) configuration of
arrows for the interacting four-vertex model with $s_1=2$ and $s_2=1$. In this
example the  arrows have an  effective size $s_1+s_2+1=4$. 
In the particular case $s_1=s_2=0$  the arrows have a unite size as it should 
be expected in the standard six-vertex model. 


Similarly as the connection of the six-vertex model with the XXZ quantum chain\cite{baxter}
we are going to show in Sec. \ref{section4} that our interacting model, with 
general parameters 
$s_1$ and $s_2$, is also related to quantum spin chains with hard-core exclusion 
interactions\cite{alcbar1,maquire}. The related models are the XXZ quantum
chains with exclusion of up spins ($\sigma^z$-basis) at distances smaller than $s_1+s_2+1$ 
\cite{alcbar1,maquire} and the 
time-evolution operator of the asymmetric exclusion problem of particles of 
size $s_1+s_2+1$, in units of lattice 
spacing \cite{alcbar5,alclazo4}.

\section{The row-to-row transfer matrix of the interacting vertex models} \label{section4}

We denote a given configuration of $n$ vertical arrows located at 
$x_1,\ldots,x_n$ along the horizontal line of the lattice by $ 
\ket{x_1,\ldots,x_n}$. The row-to-row transfer matrix $T_{s_1,s_2}$, with 
elements
\beq \label{e3}
\bra{x_1,\ldots,x_n}T_{s_1,s_2}\ket{x_1',\ldots,x_n'} = 
a_0^{n_1}b_1^{n_3}c_1^{n_5}c_2^{n_6}z_1^{n_{z_1}}z_2^{n_{z_2}},
\eeq
gives the Boltzmann weight contribution to the partition function due to 
vertical arrow configurations $\ket{x_1,\ldots,x_n}$ and $\ket{x_1',\ldots,x_n'}$ 
on two 
consecutive rows\footnote{In the particular case where we have no vertical 
arrows ($n=0$) or in the case where $s_1=s_2=0$ (six-vertex model) and $x_i=x_i'$ ($i=1,\ldots,n$) there exist two possibilities to connect the vertical arrows. 
In this case 
Eq.~\rf{e3} has two terms and should be replaced by $\bra{x_1,\ldots,x_n}T_{s_1,s_2}
\ket{x_1,\ldots,x_n} = a_1^nb_1^{L-n}+b_2a_0^{L-n}$.}. 
In \rf{e3} $n_i$ are the  numbers of vertices of type $i$ ($i=1,3,5,6$) and 
$n_{z_1}$ ($n_{z_2}$) is the number of pairs $c_1-c_2$ ($c_2-c_1$) at distance of 
$s_1$ ($s_2$) units along the horizontal line (see Fig.~\ref{fig2}). The partition 
function of the interacting vertex model, due to the periodic boundary condition 
of the lattice, is given by
\beq
\label{e4}
Z=\mbox{Tr}[T_{s_1,s_2}^M].
\eeq

As we discussed in the last section in the cases where $s_1=0, s_2\neq0$ and 
$s_1\neq 0,s_2=0$ 
the four-vertex models are equivalent to interacting five-vertex models. We can 
also verify 
directly that  the 
corresponding transfer matrices $T_{0,s_2}$ and $T_{s_1,0}$ are related to those of models with 
$s_1\neq0$ and $s_2\neq0$, namely,
\beq \label{e6}
T_{s_1,s_2} = \left( \frac{b_1}{a_0}\right)^{ns_1} {\cal P}^{-s_1} T_{0,s_1+s_2} = 
\left( \frac{a_0}{b_1}\right)^{ns_2} {\cal P}^{s_2} T_{s_1+s_2,0},
\eeq
where $\cal P$ is the horizontal translation operator, i. e., 
${\cal P}\ket{x_1,\ldots,x_n} = \ket{x_1+1,\ldots,x_n+1}$. This last relation enable us to relate 
transfer matrices of distinct models ($s_1,s_2$) and ($s_1',s_2'$), provide 
$s_1+s_2=s_1'+s_2'$, i. e., 

\beq \label{e7}
T_{s_1,s_2} = \left( \frac{b_1}{a_0}\right)^{n(s_1-s_1')} {\cal P}^{s_1'-s_1} T_{s_1',s_2'} = 
\left( \frac{a_0}{b_1}\right)^{n(s_2-s_2')} {\cal P}^{s_2-s_2'} T_{s_1',s_2'}.
\eeq


\section{The Matrix Product Ansatz for the interacting vertex models}
\label{section3}

The transfer matrix \rf{e3} with toroidal boundary conditions has a $U(1)
\times Z_L$ symmetry due to the conservation of vertical arrows 
along the horizontal lines and the translation symmetry along the 
horizontal direction. 
Consequently the associated 
Hilbert space   can be separated into block disjoint
sectors labeled by the number $n$ of vertical arrows 
($n=0,1,...,[L/(s_1+s_2+1)]$) (we denote by $[x]$ the integer part of $x$) and the
momentum eigenvalues $P$ ($P=\frac{2\pi}{L}l,l=0,1,...,L-1$) 
of the translation operator ${\cal P}=e^{-i\hat{P}}$.  We want to
solve, in each of these sectors, the eigenvalue equation
\beq
\lb{c0}
T_{s_1,s_2}|\Psi_{n,P}\rangle=\Lambda_n|\Psi_{n,P}\rangle,
\eeq
where $\Lambda_n$ and $|\Psi_{n,P}\rangle$ are the eigenvalues and
eigenvectors of $T_{s_1,s_2}$, respectively. Theses eigenvectors can be
written in general as
\beq
\lb{c0a}
\ket{\Psi_{n,P}}=\sum_{x_1,\ldots,x_n}^{*} {\cal A}(x_1,...,x_n) 
\ket{x_1,\ldots,x_n},
\eeq
where ${\cal A}(x_1,...,x_n)$ is the amplitude corresponding to the arrows
configuration with  $n$ vertical arrows  located at sites ($x_1,...,x_n$),
respectively. The symbol ($*$) in the sum means the restriction to the
configurations where 
$x_{i+1} -x_i>s_1+s_2$ ($i=1,\ldots,n-1$), $x_1 \geq 1$, 
$s_1-s_2<x_n-x_1<L-(s_1+s_2)$.
 Since $|\Psi_{n,P}\rangle$ is also an eigenvector with momentum
$P$, then the amplitudes should satisfy
\beq
\lb{c0b}
\frac{{\cal A}(x_1,...,x_n)}{{\cal A}(x_1+1,...,x_n+1)}=e^{-iP}
\eeq
for $x_n<L$, while for $x_n=L$ we have
\beq
\lb{c0c}
\frac{{\cal A}(x_1,...,L)}{{\cal A}(1,x_1+1,...,x_{n-1}+1)}=e^{-iP}.
\eeq

The exact solution of the transfer matrix \rf{e3} is obtained by an
appropriate ansatz for the unknown amplitudes ${\cal A}(x_1,...,x_n)$. As shown in
the last section our model reduces to the standard asymmetric six-vertex model
in  the case where $s_1=s_2=0$. In this particular case the model  is known to be 
exact solvable through the
 Bethe ansatz \cite{lieb,Nolden,BukmanShore} and also by the  MPA
\cite{lazo} introduced in \cite{alclazo1}-\cite{alclazo3}.
In the present paper we are going to derive the exact solution of the 
interacting vertex models with $s_1,s_2=0,1,\ldots$ by using the PMA we propose in \cite{alclazo3}.  
This ansatz asserts that to
 any amplitude in \rf{c0a} 
there exists a one-to-one correspondence with an ordered  matrix product  
\beq 
\label{c1}
{\cal A}(x_1,...,x_n) \Longleftrightarrow  E^{x_1-1}AE^{x_2-x_1-1}A \cdots E^{x_n-x_{n-1}-1} A E^{L-x_n },
\eeq
where the matrices $A$ and $E$ are associated with the sites where we have a
vertical arrow or not, respectively. Actually the objects $A$ and $E$ are 
not necessarily matrices but 
abstract operators with an associative product\footnote{
In the original formulation \cite{alclazo1,alclazo2} of the ansatz an unnecessary additional 
matrix \cite{mallick} and a trace operation was taken in the right hand side of \rf{c1}. 
The MPA  in \rf{c1} is formulated  as in the general formulation given in \cite{alclazo3}. }. Their
  commutation relations
will be fixed by imposing the validity of the eigenvalue equation of the
transfer matrix \rf{c0}. A well-defined eigenfunction is obtained, apart from
a normalization factor, if all the amplitudes are uniquely
related. Equivalently, in the subset of words (products of matrices) of the
algebra containing $n$ matrices $A$ and ($L-n$) matrices $E$, should  exist
only a single independent word. The relation between any two words gives a 
$c$ number that is equal to the
ratio between the corresponding amplitudes in \rf{c1}.

Since the eigenfunctions $|\Psi_{n,P}\rangle$ \rf{c0a} have a well-defined
momentum $P=\frac{2\pi}{L}j$ ($j=0,...,L-1$), the relations \rf{c0b} and
\rf{c0c} imply the following constraints for the matrix products appearing in
the ansatz \rf{c1}
\bea
\lb{c2a}
E^{x_1-1}A&&E^{x_2-x_1-1}A \cdots E^{x_n-x_{n-1}-1} A E^{L-x_n }=
\nonumber \\
&&\e^{-iP}E^{x_1}AE^{x_2-x_1-1}A \cdots E^{x_n-x_{n-1}-1} A E^{L-x_n-1},
\eea
for $x_n<L$, and for $x_n=L$
\bea
\lb{c2b}
E^{x_1-1}A&&E^{x_2-x_1-1}A \cdots E^{x_n-x_{n-1}-1} A =
\nonumber \\ 
&&\e^{-iP} A E^{x_1}AE^{x_2-x_1-1}A \cdots E^{x_n-x_{n-1}-1} A E^{L-x_{n-1}-1}.
\eea

In order to proceed, in the usual way, we are going to consider first the
eigensectors of $T_{s_1,s_2}$ with small values of $n$.

{\bf The case n = 0.}

In this case the solution of the eigenvalue equation \rf{c0} is trivial
since we do not have vertical arrows between two successive rows. There are
only two possible horizontal arrangements:  either all
bonds have a horizontal arrow or all of them are empty. In this case the
vertices are all of type $1$ or type $3$ (see figure 1) and consequently the
eigenvalue is given by  
\beq
\label{c4}
\Lambda_0=a_0^L+b_1^L.
\eeq

{\bf The case n = 1.}

We have in this case just one vertical arrow between two rows. The
transfer matrix links a vertical solid line at position $x$ ($x=1,\ldots,L$)
above a row to a vertical line at any position $y$ ($L\ge x \ge y+s_1$ or
$y-s_2\ge x\ge 1$) under this row. The elements of the transfer matrix
$T_{s_1,s_2}(y,x)$ in this sector with momentum $P$ are given by \rf{e3}. They
are given by the product of the Boltzmann weights of the vertex appearing on the
row. The vertex configuration at the sites $x$ and $y$ will be of types $c_2$
and $c_1$. If  the position of the line $x$ is greater or equal (less or equal)
 than $y$ and all the
others vertices will be of types $b_1$ ($a_0$) and $a_0$ ($b_1$) depending on
whether the vertices are between the positions $x$ and $y$, or not,
respectively. 
 However in the special cases where $x=y+s_1$ or $x=y-s_2$   we have also 
to include the fugacities $z_1$ ($z_2$) due to the   long range interaction
 \rf{e2} between these vertices. Consequently the eigenvalue equation
\rf{c0} for the transfer matrix \rf{e3} 
associated with the components of
$|\psi_{n,P}\rangle$ \rf{c1} with $n=1$ and momentum $P$, give us the 
algebraic relations
for the matrix products
\bea
\label{c5}
&&\Lambda_1 E^{x-1}AE^{L-x} = \sum_{y=1}^{x-s_1-1}a_0^{L-x+y-1}b_1^{x-y-1}c_1c_2 E^{y-1}A E^{L-y} \nonumber\\
&& +\sum_{y=x+s_2+1}^{L}a_0^{y-x-1}b_1^{L-y+x-1}c_1c_2 E^{y-1}AE^{L-y}\nonumber\\
&& +a_0^{L-s_1-1}b_1^{s_1}b_2 E^{x-s_1-1}AE^{L-x+s_1} +b_1^{L-s_2-1}a_0^{s_2}a_1 E^{x+s_2-1}AE^{L-x-s_2},
\eea
where we inserted \rf{e2} in the last two terms in the right hand side.
 
The equation \rf{c5} is simplified by expressing  all the matrix
products in terms of a single one. This is done 
 by imposing that the matrix $A$ depend on a
spectral parameter $k$. Without loss of generality, the matrix $A$ can be
written in terms of the matrix $E$ and a spectral parameter dependent matrix
$A_k$\footnote{
 The most general relation $A=\sum_{j=1}^n
  A^{\alpha}A_{k_j}E^{\beta}$ could be used. However \rf{c6} is more
  convenient since otherwise the structure constants $S(k_j,k_l)$ that 
will appear in \rf{c25}
  will depend on the parameters $s_1$ and $s_2$.  }

\beq
\label{c6}
A=E^{-s_1}A_kE^{1-s_2},
\eeq
with $A_k$ satisfying the following algebraic relation with the matrix $E$
\beq
\label{c7}
EA_k=\e^{ik}A_kE.
\eeq
Inserting \rf{c6} in \rf{c5} and using \rf{c7} we can factorize \rf{c5}:
\bea
\label{c8}
&&\{ \Lambda_1 - \sum_{y=1}^{x-s_1 -1}a_0^{L-x+y-1}b_1^{x-y-1}c_1c_2 
e^{ik(y-x)}- 
\sum_{y=x+s_2+1}^{L}a_0^{y-x-1}b_1^{L-y+x-1}\nonumber \\
&&\times c_1c_2 \e^{ik(y-x)} 
 - a_0^{L-s_1-1}b_1^{s_1}b_2\e^{-iks_1}\nonumber \\
&&-b_1^{L-s_2-1}a_0^{s_2}a_1\e^{iks_2}\}E^{x-s_1-1}A_kE^{L-x+1-s_2}=0.
\eea
In order to produce a nonzero norm state we should impose
$E^{x-s_1-1}A_kE^{L-x+1-s_2}\ne 0$, for $x=1,...,L$. As a consequence we must
have, by evaluating the sums in \rf{c8},
\bea 
\label{c9}
\Lambda_1 = a_0^L 
&&L(k)\left(\frac{b_1}{a_0}\right) 
^{s_1}\e^{-iks_1} + 
b_1^L M(k)\left(\frac{a_0}{b_1} \right)^{s_2}\e^{iks_2}  \nonumber \\ 
&& + a_0^L \frac{c_1c_2}{b_1}\left( \frac{b_1}{a_0}
\right)^{x} \frac{\e^{ik(1-x)}}{b_1-a_0\e^{ik}}(1 - \e^{ikL}),
\eea
where
\beq 
\label{c10}
L(k) = \frac{a_0b_2+(c_1c_2-b_1b_2)\e^{-ik}}{a_0^2-a_0b_1\e^{-ik}}, 
\quad M(k) = \frac{a_0a_1-c_1c_2-a_1b_1i\e^{-ik}}
{a_0b_1-b_1^2\e^{-ik}}.
\eeq 
In order to satisfy \rf{c5}, the eigenvalue $\Lambda_1$ in \rf{c9} should
independ on the vertical line position
$x$. Thus the last term in the right hand side of \rf{c9} must vanish. The
only way to cancel this term, for non zero Boltzmann weights, is obtained 
by imposing $e^{ikL} =1$, that fixes  
 the spectral parameter $k$
\beq
\label{c11} 
 k=\frac{2\pi}{L}j \;\;\;\;\; (j=0,1,\ldots,L-1).
\eeq 
 The
eigenvalue \rf{c9} is then given by
\beq 
\label{c12}
\Lambda_1 = a_0^L L(k)\left(\frac{b_1}{a_0} \right)^{s_1}\e^{-iks_1} + b_1^L M(k)\left(\frac{a_0}{b_1} \right)^{s_2}\e^{iks_2},
\eeq
where the values of the spectral parameter $k$ are given by \rf{c11}.

Finally, by inserting \rf{c6} and using \rf{c7} in the equations \rf{c2a} and
\rf{c2b} we verify that $k$ coincides with the momentum of the eigenstate:
\beq
k=P=\frac{2\pi}{L}j \;\;\; (j=0,1,...,L-1).
\eeq

{\bf The case n = 2.}

In this sector there are two vertical arrows in the row. We have in general two
types of relations for the amplitudes \rf{c1}. Relations where at
least one of the vertical arrows ($y_1$,$y_2$) are at distance $s_1$ or $s_2$
from ($x_1$,$x_2$), and those where $y_1$ and $y_2$ interlace with $x_1$ 
and $x_2$:
($1\leq y_1<x_1-s_1$; $x_1+s_2<y_2<x_2-s_1$) and ($x_1+s_2<y_1<x_2-s_1$; 
$x_2+s_2<y_2\leq L$).
 Then, the eigenvalue equation \rf{c0}
imply 
\bea 
\label{c18}
&&\Lambda_2  E^{x_1-1}AE^{x_2-x_1-1}AE^{L-x_2} = 
 \sum_{y_1=1}^{x_1-s_1}\sum_{y_2=x_1+s_2}^{x_2-s_1\;\;*} 
a_0^{L-(x_2-y_1+1)}c_2 \nonumber \\
&& \times g(x_1,y_1)f(y_2,x_1)g(x_2,y_2) E^{y_1-1}AE^{y_2-y_1-1}A
E^{L-y_2} \nonumber \\
&&+  \sum_{y_1=x_1+s_2}^{x_2-s_1}\sum_{y_2=x_2+s_2}^{\;\;\;\;L\;\;*}
 b_1^{L-(y_2-x_1+1)}c_1f(y_1,x_1) \nonumber \\
&& \times g(x_2,y_1)f(y_2,x_2) E^{y_1-1}AE^{y_2-y_1-1}A
E^{L-y_2},
\eea
where the symbol $*$ in the sums means that terms with $|y_2-y_1|\le s_1+s_2$ are excluded
and 
\beq
\label{c19}
f(y,x)=\left\{ \begin{array}{cc}
              c_2 a_0^{y-x-1}  & \mbox{if}\;\;y>x+s_2 \\
              a_0^{s_2}a_1/c_1 & \mbox{if}\;\;y=x+s_2
              \end{array}
       \right.,
\eeq
\beq \label{c19p}
g(x,y)=\left\{ \begin{array}{cc}
                     b_1^{x-y-1}c_1  & \mbox{if}\;\;x>y+s_1 \\
              b_1^{s_1}b_2/c_2  & \mbox{if}\;\;x=y+s_1
              \end{array}
       \right..
\eeq
The relation \rf{c18} connects configurations where the arrangements of vertical
arrows ($x_1,x_2$) above the row do not have the same distance of the vertical
arrows ($y_1,y_2$) bellow the  row. In other words, the distance of the incoming arrows $y_2-y_1$ are in
general different of the outcoming distance $x_2-x_1$. To solve the relation
\rf{c18} we need now to use a generalization of the algebraic relation \rf{c6}
for the case of two arrows. The generalization of \rf{c6} is done by writing
the matrix $A$ in terms of two new spectral parameter matrices as:
\beq   
\label{c20}                         
A=\sum_{j=1}^2 E^{-s_1}A_{k_j}E^{1-s_2},
\eeq
with the commutation relations 
\beq
\label{c21}
EA_{k_j}=\e^{ik_j}A_{k_j}E \;\;\;\;\; (j=1,2),
\eeq
where the spectral parameters $k_1$ and $k_2$ are up to now unknown complex
numbers.

Inserting \rf{c20} in \rf{c18} and using  \rf{c21} and
\rf{c10} we obtain, after similar manipulation as we did in the case $n=1$,
the following constraint

\bea
\label{c22}
&&\sum_{j,l=1}^2\left[ \Lambda_2
  -a_0^L\left(\frac{b_1}{a_0}\right)^{2s_1}\e^{-i(k_j+k_l)s_1}
  L(k_j)L(k_l)
-b_1^L\left(\frac{a_0}{b_1} \right)^{2s_2}\e^{i(k_j+k_l)s_2}\right.\nonumber \\
&&  \times M(k_j)M(k_l)\left. \right]
  \e^{ik_js_1}\e^{-ik_ls_2} \e^{ik_jx_1}\e^{ik_lx_2}A_{k_j}A_{k_l}\nonumber \\
&&-\sum_{j,l=1}^2 a_0^L\left[ L(k_l)M(k_j)-\frac{a_1b_2}{a_0b_1} \right]\left( 
\frac{b_1}{a_0}\right)^{s_1-s_2} \left( \frac{b_1}{a_0}
\right)^{x_2-x_1}\e^{i(k_j+k_l)(s_1-s_2)} \nonumber \\
&& \times \e^{i(k_j+k_l)x_1} A_{k_j}A_{k_l}  -\sum_{j,l=1}^2 b_1^L\left[ L(k_l)M(k_
j)-\frac{a_1b_2}{a_0b_1} \right]\left( \frac{b_1}{a_0}\right)^{s_1-s_2}\left(
  \frac{b_1}{a_0} \right)^{x_1-x_2}\nonumber \\
&&\times \e^{i(k_j+k_l)(s_1-s_2)}
 \e^{i(k_j+k_l)x_2} A_{k_j}A_{k_l}\nonumber \\
&&+\sum_{j,l=1}^2
a_0^L\frac{c_1^2c_2^2\left[\e^{iLk_l}\e^{ik_j(s_2+s_1)}\e^{-ik_ls_2}\e^{ik_jx_1}-\e
^{ik_js_1}\e^{ik_lx_1} \right]
  }{b_1^2(a_0-b_1\e^{-ik_j})(a_0-b_1\e^{-ik_l})} \nonumber \\
&& \times \left( \frac{b_1}{a_0} \right)^{-s_2}\left( \frac{b_1}{a_0} \right)^{x_2}
 A_{k_j}A_{k_l} \nonumber
\\
&&-\sum_{j,l=1}^2 a_0^L\left[\frac{c_1^2c_2^2\left[\e^{iLk_l}\e^{-ik_ls_2}\e^{-ik_j
}\e^{ik_jx_2}-\e^{ik_js_1}\e^{-ik_l(s_2+s_1)}\e^{-ik_l}\e^{ik_lx_2} \right] }{a_0b_
1(a_0-b_1\e^{-ik_j})(a_0-b_1\e^{-ik_l})}\right]\nonumber \\
&& \times \left( \frac{b_1}{a_0}\right)^{s_1}\left( \frac{b_1}{a_0} \right)^{x_1} A
_{k_j}A_{k_l}
\nonumber \\
&&+\sum_{j,l=1}^2 a_0^Lc_1c_2a_1\left[\frac{\e^{iLk_l}\e^{ik_j(s_2+s_1)}\e^{-ik_ls_
2}\e^{ik_jx_1}
  }{b_1^2(a_0-b_1\e^{-ik_l})}-
  \frac{\e^{ik_js_1}\e^{ik_lx_1}}{b_1^2(a_0-b_1e^{-ik_j})}\right]\nonumber \\
&&\times \left( \frac{b_1}{a_0} \right)^{-s_2} \left( \frac{b_1}{a_0} \right)^{x_2}
 A_{k_j}A_{k_l} \nonumber
\\
&&+\sum_{j,l=1}^2 a_0^Lc_1c_2b_2\left[\frac{\e^{iLk_l}\e^{-ik_ls_2}\e^{ik_jx_2}
  }{a_0b_1(a_0-b_1\e^{-ik_l})}-
  \frac{\e^{ik_js_1}\e^{-ik_l(s_2+s_1)}\e^{ik_lx_2}}{a_0b_1(a_0-b_1\e^{-ik_j})}
\right] \nonumber \\
&&\times \left( \frac{b_1}{a_0} \right)^{s_1} \left(
  \frac{b_1}{a_0} \right)^{x_1} A_{k_j}A_{k_l} = 0,
\eea            
where $1\leq x_1 < x_2 \leq L$ with $|x_2-x_1|\ge s_1+s_2+1$. This can
only be satisfied if each sum is identically zero. Moreover since $\Lambda_2$
should be independent of $x_1$ or $x_2$ a possible solution of  \rf{c22} is
obtained by imposing
\bea
\label{c23}
\Lambda_2=&&a_0^L\left(\frac{b_1}{a_0}\right)^{2s_1}\e^{-i(k_1+k_2)s_1}L(k_1)L(k_2)+b_1^L\left(\frac{a_0}{b_1} \right)^{2s_2}\e^{i(k_1+k_2)s_2}
 \nonumber \\ 
&& \times M(k_1)M(k_2).
\eea
The algebraic relation between the matrices $A_{k_1}$ and $A_{k_2}$ are obtained
by imposing that both the second and third terms in \rf{c22} are zero independently,
i. e., 
\beq
\label{c24}
A_{k_j}A_{k_l}=S(k_j,k_l)A_{k_l}A_{k_j}, \;\;\;\;\ \left(
  A_{k_j} \right)^2=0 \;\;\;\;\; (j\neq l=1,2),
\eeq
where
\beq
\label{c25}
S(k_j,k_l)=-\frac{L(k_j)M(k_l)-\frac{a_1b_2}{a_0b_1}}{L(k_l)M(k_j)-\frac{a_1b_2}{a_0b_1}},
\eeq
with $L(k)$ and $M(k)$ given by \rf{c10}. Note that the structure constants
$S(k_j,k_l)$ is independent of $s_1$ and $s_2$ due to the choice
\rf{c20} (see the footnote related to Eq.~\rf{c6}).
 Finally, the vanishing of the last four terms in \rf{c22} will give
us relations that fix the spectral parameters values $k_1$ and $k_2$. These
equations are obtained by exploring the algebraic relations 
\rf{c24} 
\beq
\label{c26}
\e^{iLk_j}=\left(\frac{\e^{ik_j}}{\e^{ik_l}}\right)^{s_1+s_2}S(k_j,k_l) \;\;\;\;\; (l,j=1,2 \;\; \mbox{and} \;\; l\neq j).
\eeq
The eigenvalues  are obtained by inserting the solutions
($k_1,k_2$) of these last equations in \rf{c23}. 
 The momentum $P$ is obtained by using \rf{c20} and \rf{c21} in
\rf{c2a} and \rf{c2b}: $P=k_1+k_2$.
 The amplitudes of the 
corresponding eigenvectors are obtained from the  algebraic relations \rf{c20},
\rf{c21} and \rf{c24}.

{\bf{The case of general n.}}

The general case follows straightforwardly from the $n=2$ case. The previous
calculation can be extended for arbitrary values of the number $n$ of vertical
arrows. The eigenvalue equation for the transfer matrix connects the 
amplitudes ${\cal A}(x_1,\ldots,x_n)$ and ${\cal A}(y_1,\ldots,y_n)$ where:
$1\leq y_1 \leq x_1-s_1, x_1+s_2\leq y_2 \leq x_2-s_1, \ldots, 
x_{n-1} +s_2 \leq y_n\leq x_n -s_1$ and 
$x_1+s_2 \leq y_1 \leq x_2-s_1, x_2+s_2\leq y_2 \leq x_3-s_1, \ldots, 
x_n-s_2 \leq y_n \leq L$.
To solve the eigenvalue equation we need to extend the definition
\rf{c20} and the commutation relations \rf{c21} for general $n$, i. e.,
\beq
\label{c28}
A=\sum_{j=1}^n E^{-s_1}A_{k_j}E^{1-s_2}
\eeq
with
\beq
\label{c29}
EA_{k_j}=\e^{ik_j}A_{k_j}E \;\;\;\;\; (j=1,...,n).
\eeq
The parameters  $k_j $ ($j =1,\ldots,n$) are in general unknown complex numbers that
will be fixed by the eigenvalue equation \rf{c0}. Inserting \rf{c28} in 
 \rf{c0} and using the commutation relations \rf{c29} we
obtain, similarly as done in the case $n=2$, the algebraic relations among the
matrices $A_{k_j}$ ($j=1,\ldots,n$):
\beq
\label{c30}
A_{k_j}A_{k_l}=S(k_j,k_l)A_{k_l}A_{k_j} \;\;\;\;\; \left(
  A_{k_j} \right)^2=0 \;\;\;\;\; (j\neq l=1,\dots,n),
\eeq
where $S(k_j,k_l)$ is given by \rf{c25} and the spectral parameters $k_j$
  ($j=1,...,n$) are fixed by the equation
\beq
\label{c31}
e^{iLk_j}=-\prod_{l=1}^n
\left(\frac{\e^{ik_j}}{\e^{ik_l}}\right)^{s_1+s_2} S(k_j,k_l) \;\;\;\;\; (j=1,...,n).
\eeq
The acceptable set $\{k_j\}$ of spectral parameters defining the
eigenvectors $|\Psi_{n,P}\rangle$ is given by 
 the solutions of \rf{c31} where $k_l\neq
k_j$ ($j,l=1,...,n$). Since $\left(A_{k_j} \right)^2=0$, solutions of \rf{c31}
with coinciding roots give us null states. 

 No new algebraic relations appear for the matrices $\{A_{k_j}\}$ besides
\rf{c29} and \rf{c30}. This not easy to verify. The difficulty here is similar to
that in showing the absence of many-particle scattering in the standard
coordinate Bethe ansatz for vertex models \cite{lieb}-\cite{gaudin}.
The product associativity of the algebra \rf{c29}
and \rf{c30} follows from the property $S(k_j,k_l)S(k_l,k_j)=1$. From the the
eigenvalue equation \rf{c0} we  obtain the eigenvalues 
\beq
\label{c32}
\Lambda_n = a_0^L\left( \frac{b_1}{a_0} \right)^{ns_1}\prod_{j=1}^n 
\e^{-ik_js_1}L(k_j)+ b_1^L\left( \frac{a_0}{b_1} \right)^{ns_2}\prod_{j=1}^n \e^{ik_js_2}M(k_j),
\eeq 
where $L(k)$ and $M(k)$ are given by \rf{c10} and the spectral parameters $\{k_j \}$
are the solutions of \rf{c31}. 
Finally, the momentum $P$ follows from \rf{c2a},
\rf{c2b}, \rf{c28} and \rf{c29}:
\beq
\label{c33}
P=\sum_{j=1}^n k_j.
\eeq

Before closing this section let us give a possible representation for the
matrices $E$ and $A$ of the ansatz \rf{c1} \cite{alclazo3}. 
For a given solution
$\{k_1,...,k_n \}$ of the spectral parameter equations \rf{c31}, in the sector
with $n$ vertical arrows, the matrices $E$ and $\{A_{k_1},...,A_{k_n} \}$ have
the following $2^n\times 2^n$ dimensional representation:
\beq
\lb{c34}
E=\bigotimes_{l=1}^n
\left( \begin{array}{cc}
1  &   0  \\
0  &  \e^{-ik_l} 
\end{array} \right),
\eeq
\beq
\lb{18}
A_{k_j}=\left[ \bigotimes_{l=1}^{j-1}
\left( \begin{array}{cc}
S(k_j,k_l)  &   0  \\
0  &  1 
\end{array} \right)\right]\bigotimes
\left( \begin{array}{cc}
0  &   1  \\
0  &   0 
\end{array} \right)\bigotimes_{l=j+1}^n
\left( \begin{array}{cc}
1  &   0  \\
0  &   1 
\end{array} \right),
\eeq
where $S(k_j,k_l)$ are given by \rf{c25} and $A$ is obtained from
\rf{c28}. 
 It is important to notice that the
matrices $E$ and $A_{k_j}$ have the same functional form obtained for the
formulation of the MPA for the Hamiltonian of the XXZ spin chain
\cite{alclazo3}. It happens because the matrices $E$ and $A_{k}$ satisfy the
same algebraic relations \rf{c28}-\rf{c30} in both models. The only
difference are the values of the spectral parameters $\{k_1,...,k_n\}$ that
for the present model are fixed by equation \rf{c31}. 

It is important to notice  that all the interacting four-vertex 
($s_1\neq 0,s_2\neq 0$), five-vertex ($s_1=0,s_2\neq0$ and $s_1\neq0,s_2=0$) 
and six-vertex models ($s_1=s_2=0$)  have the same functional form   for the 
operators $A_{k_j}$ ($j=1,\ldots,n$). The only difference being the structure 
constants   
 $S(k_j,k_l)$, since the spectral parameters $k_1,\ldots,k_n$ are solutions 
of distinct equations.  As a test of our calculations we see that 
 \rf{c31}, 
for the case $s_1=s_2=0$ coincides with the spectral parameter equations 
obtained by the well known Bethe ansatz calculation of the asymmetric 
six-vertex model  
\cite{Nolden,BukmanShore}. Using the representation \rf{c34} and \rf{18} 
we also obtain on this case the same amplitudes for the eigenfunctions as 
those derived by the Bethe ansatz.

 We can also solve the interacting models introduced in this paper by
the use of the same coordinate Bethe ansatz used in the six-vertex model
\cite{lieb}-\cite{gaudin}. 
In this case instead of obtaining the structure constants \rf{c25}, 
with no dependence on $s_1$ or $s_2$, we obtain a S-matrix that depends on these
parameters. Although the final spectral parameter equations \rf{c31} and the
eigenfunctions for both ansatz are the same,  we believe the derivation
through our MPA is more elegant.

The phase diagram of the interacting vertex models introduced in this
paper can be derived from that of the six-vertex model. This follows from the
spectral-parameter equations \rf{c31}, that can be written as
\beq \label{c35}
\e^{i[L-n(s_1+s_2)]k_j} \e^{iP(s_1+s_2)} = -\prod_{l=1}^n S(k_j,k_l), \quad (j=1,\ldots,n),
\eeq
where we have used \rf{c33}. Consequently the eigenvalues belonging
to the eigensector with $n$ vertical arrows and momentum $P$ of
the transfer matrix $T_{s_1,s_2}$ with parameters $s_1$ and $s_2$ are related
to those of the standard six-vertex model ($s_1=s_2=0$). The related
six-vertex model is defined on a cylinder with perimeter  $L' = L- n(s_1+s_2)$ and
with  a momentum-dependent seam spanning its length\footnote{ The phase $\e^{iP(s_1+s_2)}$ in the left hand side
of \rf{c35} can be obtained by considering a vertex model with twisted boundaries. The twisted is obtained by
 introducing  a seam  with distinct vertex fugacities along the vertical
 direction.}.

\section{The interacting vertex models and generalized XXZ quantum chains}
\label{section5}

It is well known the connections among the XXZ quantum chain and the standard 
asymmetric six-vertex model \cite{baxter}. The row-to-row transfer matrix 
of the six-vertex model is the generating function of an infinite set of 
commuting charges. The XXZ quantum Hamiltonian being one of these charges. 
Since the six-vertex model is a particular case ($s_1=s_2=0$) of our general 
model it is natural to expect that we can also generate generalized 
exact solvable  XXZ quantum chains for the general case where $s_1,s_2 \geq0$.

The first step towards the derivation of those generalized quantum chains is 
the identification of the row-to-row transfer matrix $T_{s_1,s_2}$, for 
arbitrary values of $s_1$ and $s_2$, as a generating function of commuting 
charges. 

It is important to notice that the structure constants $S(k_j,k_l)$ given in 
\rf{c25} do not depend on the particular values of $s_1$ and $s_2$. 
 It only depends on the  parameters $a_0$, $a_1$, $b_1$, 
$b_2$, $c_1$ and $c_2$. Consequently it is convenient to use the same 
parametrization used in \cite{dahmen} for the asymmetric 
 six-vertex model
\bea
\lb{r1}
&& a_0 = \e^{-\beta \varepsilon_1} = \e^{h+\nu}\frac{\sinh
  (\gamma+\epsilon\mu)}{\sinh \gamma} \nonumber \\
&& a_1 = \e^{-\beta \varepsilon_2} = \e^{-h-\nu}\frac{\sinh
  (\gamma+\epsilon\mu)}{\sinh \gamma} \nonumber \\
&& b_1 = \e^{-\beta \varepsilon_3} = \e^{-h+\nu}\frac{\sinh
  \mu}{\sinh \gamma } \\
&& b_2 = \e^{-\beta \varepsilon_4} = \e^{h-\nu}\frac{\sinh
  \mu}{\sinh \gamma } \nonumber \\
&& c_1 = \e^{-\beta \varepsilon_5} = 1 \nonumber \\
&& c_2 = \e^{-\beta \varepsilon_6} = 1 \nonumber,
\eea
where $\epsilon =\pm1$ and $\gamma$, $\mu$, $h$ and $\nu$ are free 
parameters. The choice $c_1=c_2=1$ does not restrict the model since, 
due to the periodic boundary condition, these parameters always appear 
in pairs. The symmetric interacting models are obtained by setting 
$h=\nu=0$. 

Inserting in \rf{c25} the parameters in \rf{r1} we obtain 
\beq \label{r2}
S(k_j,k_l) = - \frac
{ 1 +\e^{4h}\e^{i(k_j+k_l)} -\epsilon \e^{2h}\cos {\gamma} \e^{ik_l}}
{ 1 +\e^{4h}\e^{i(k_j+k_l)} -\epsilon \e^{2h}\cos {\gamma} \e^{ik_j}},
\eeq
that shows the independence of the structure constants of the algebra 
defining the MPA on the parameters $\mu$ and $\nu$.

The results of previous sections imply that, for a given solution 
$\{k_1,\ldots,k_n\}$ of the spectral parameter equations \rf{c31}, the amplitudes 
of the corresponding eigenfunction of $T_{s_1,s_2}$ depends only on 
the algebraic relations \rf{c28}-\rf{c30}. Consequently all the 
eigenfunctions of $T_{s_1,s_2}$, for fixed values of $s_1$ and $s_2$, are 
the same for arbitrary values of $\mu$ and $\nu$, i. e., 
\beq \label{r3}
[T_{s_1,s_2}(\mu,\nu), T_{s_1,s_2}(\mu',\nu')] = 0.
\eeq
Moreover, since from \rf{e6} $T_{s_1,s_2}$ and $T_{s_1',s_2'}$ differ 
by the multiplication of a diagonal operator, we may write the general 
relation
 
\beq \label{r4}
[T_{s_1,s_2}(\mu,\nu), T_{s_1',s_2'}(\mu',\nu')] = 0, \;\;\;\; 
s_1+s_2=s_1'+s_2'.
\eeq

The conserved charges are derived by expanding $T_{s_1,s_2}(\mu,\nu)$ 
in terms of $\mu$.
Since we are interested only in the derivation of the quantum chain 
associated to $T_{s_1,s_2}(\mu,\nu)$, in the following we are going to 
consider only the two leading terms in the $\mu$-expansion.

The matrix elements of $T_{s_1,s_2}(\mu,\nu)$ in the eigensector with 
$n$ arrows are given by\footnote{ The exceptional cases considered in the 
footnote of page $7$ should be considered separately, with 
no changes in the final results.}
\bea \label{a1}
\bra{\{y\}} &&T_{s_1,s_2}(\mu,\nu) \ket{\{x\}} = P_s(\{y\}) P_s(\{x\}) 
a_0^{L-(x_n-y_1+1)} c_2  \nonumber \\
 && \times g(x_1,y_1)f(y_2,x_1)g(x_2,y_2)\cdots f(y_n,x_{n-1}) 
g(x_n,y_n),
\eea
if $x_1 \geq y_1+s_1$, and 
\bea\label{a2}
\bra{\{y\}} &&T_{s_1,s_2}(\mu,\nu) \ket{\{x\}} = P_s(\{y\}) P_s(\{x\}) 
b_1^{L-(y_n-x_1+1)} c_2 \nonumber \\
&& f(y_1,x_1)g(x_2,y_1)f(y_2,x_2)\cdots g(x_n,y_{n-1})f(y_n,x_n), 
\eea
if $x_1 \leq y_1-s_2$. In these last equations $s= s_1+s_2$ and  
$P_s(\{x\})$ projects out configurations not satisfying the hard-core 
exclusion constraint:
\beq \label{a3}
P_s(\{x\}) = \theta (x_n-x_1 -[L-(s+1)]) 
\prod_{i=1}^{n-1}\theta(x_{i+1}-x_i -(s+1)),
\eeq
where 
\beq
\label{a3p}
\theta(y)=\left\{ \begin{array}{rc}
              1 & \mbox{if} \quad y\le 0  \\
              0 & \mbox{if} \quad y<0  
              \end{array} \right.
\eeq
is the standard step function.
In \rf{a1} and \rf{a2} the functions $f(y,x)$ and $g(x,y)$ are defined in 
\rf{c19} and \rf{c19p}  and have the leading behaviour in the $\mu$-expansion
\bea \label{a4}
f(y,x) =&& \e^{(h+\nu)(y'-x-s_1-2)} \left\{ 1 + \left[(y'-x-s_1-2)+2\delta_{y'-x,s+1}\right]
\epsilon \mu \coth \gamma\right\} \nonumber \\
&& \times \theta(y'-x-(s+1))\delta_{y,y'-s_1-1} 
+ O(\mu^2),
\eea
\bea \label{a5}
g(x,y) =&& \left(\frac{\e^{\nu-h}\mu}{\sinh{\gamma}}\right)^{s_1}
\left[\delta_{x-y',0}+ \frac{\mu}{\sinh{\gamma}}
(\e^{\nu-h} \delta_{x-y',1} + \e^{h-\nu}\delta_{x-y',-1})\right] \nonumber \\
 && \times \delta_{y,y'-s_1-1} 
+ O(\mu^{s_1+2}).
\eea
Using \rf{a4} and \rf{a5} in \rf{a1} we obtain
\bea \label{a7}
&&\bra{\{y\}} T_{s_1,s_2}\ket{\{x\}} = 
\left( \frac{\e^{\nu-h}\mu}{\sinh{\gamma}}\right)^{ns_1} \e^{h+\nu} 
P_s(\{y\}) \left\{\prod_{i=1}^{n}\delta_{y_i,y_i'-s_1 -1}\right\} \nonumber \\
&&\times \left\{ 1 + \frac{\mu}{\sinh{\gamma}} 
\sum_{i=1}^{L} [ \e^{-2h} \delta_{x_i,y_i'+1}^{(L)} +
\e^{2h}\delta_{x_i,y_i'-1}^{(L)} +
\left(L-2n-ns_1  \right. \right. \nonumber \\
&& \left. \left. \left. +2\sum_{i=1}^L \delta_{x_{i+1}-x_i,s+1}^{(L)}\right)\epsilon \cosh{\gamma}
\right] + O(\mu^2) \right\} P_s(\{x\}),
\eea
where 
\beq \label{a8}
\delta_{x_{i+1}-x_i,s+1}^{(L)} =\left\{ \begin{array}{ccl}
              \delta_{x_{i+1}-x_i,s+1}  & \mbox{for} & i=1,\ldots,n-1  \\
              \delta_{x_{n}-x_1,L-(s+1)}  & \mbox{for}&  i=n\quad  
              \end{array} \right. .
\eeq
We can rewrite \rf{a7} in terms of spin-$\frac{1}{2}$ Pauli matrices by 
identifying 
an arrow at site $i$ as $\sigma_i^z=+1$, and we obtain
\bea \label{a9}
T_{s_1,s_2}(\mu,\nu) &=& \e^{-[h(1+s_1)+\nu]\sum_{i=1}^L(\sigma_i^z+1) + L(h+\nu)} 
\left(\frac{\mu}{\sinh{\gamma}}\right)^{s_1\sum_{i=1}^L\frac{\sigma_i^z+1}{2}} 
{\cal{P}}^{-(1+s_1)} \nonumber \\
&& \times \left[ P_{s_1+s_2} + \frac{\mu}{\sinh{\gamma}} H_{s_1,s_2} + O(\mu^2)\right],
\eea
where 
\bea \label{a10}
H_{s_1,s_2} &=& P_{s_1+s_2}\left\{ \sum_{i=1}^L \left[ 
\epsilon_+\sigma_i^-\sigma_{i+1}^+ + \epsilon_- \sigma_i^+\sigma_{i+1} + 
\frac{\Delta}{2}(\sigma_i^z\sigma_{i+s_1+s_2+1}^z + 1)\right] \right. \nonumber \\ 
&&\left. -h_m\sum_{i=1}^L (1 +\sigma_i^z)\right\}P_{s_1+s_2}, \nonumber \\
\epsilon_+ &=& e^{2h}, \quad \epsilon_- = \e^{-2h}, \quad h_m=\frac{s_1}{2}, 
\quad \Delta= \epsilon \cosh{\gamma}, \quad \epsilon = \pm 1.
\eea
In this last equation periodic boundary conditions 
is imposed, $\sigma_{i}^{\pm} = (\sigma^x \pm i\sigma^y)/2$ and $P_{s_1+s_2}$ is the diagonal operator 
$\bra{\{x\}}P_{s_1+s_2}\ket{\{x\}} = P_{s_1+s_2}(\{x\})$ defined in \rf{a3}, that 
projects out configurations where two up spins are distances smaller than 
$s_1+s_2+1$. The same expressions \rf{a9} and \rf{a10} are also obtained by 
inserting \rf{a4}  and \rf{a5} in 
\rf{a2}. 

We see from \rf{a10} that associated to the family of solvable interacting 
vertex models, with interacting parameters $s_1$ and $s_2$, we have a generalized 
asymmetric XXZ quantum chain. This quantum chain have hard-core exclusion 
interactions that forbid the occupation of up spins at distances smaller 
than $s_1+s_2+1$. The important parameter controlling the exclusion effect is 
given by the combination $s=s_1+s_2$. The effect of the parameter $s_1$ in 
\rf{a10} is the same as an external magnetic field $h_m= \frac{s_1}{2}$.

The Hamiltonian \rf{a10} for the case where $s_1=s_2=0$ gives the standard 
asymmetric XXZ chain. The particular case where $\Delta = \epsilon_+   + \epsilon_-$ 
gives the time-evolution operator of the asymmetric exclusion problem where 
the particles have hard-core size $s_1+s_2+1$, in units of lattice spacing
 \cite{sazamoto,alcbar5}.

The  integrability of \rf{a10}, for general values of $s_1$ and $s_2$, 
is already known in the literature \cite{sazamoto,alcbar5,alcbar6}.
What is unknown is why the inclusion of the hard-core exclusion constraints in 
the standard XXZ quantum chain model ($s_1 =s_2=0$), producing the Hamiltonian 
\rf{a10}, does not destroy the  integrability of the original model. This 
paper gives the explanation for this fact. The quantum chains \rf{a10}, for a 
given value of $s_1$ and $s_2$, belongs to the infinite set of of commuting 
charges generated by the row-to-row transfer matrix $T_{s_1+s_2} $ of 
our interacting vertex models.

\section{ Conclusions and generalizations }\label{section6}

We have introduced in this paper a special family of exact solvable
interacting vertex models. 
These models are generalizations of the  six-vertex model.
 Besides the usual nearest-neighbor interactions
imposed by the lattice connectivity, the models also contains  hard-core
interactions along the horizontal lines of the lattice. The range of the
additional interactions depends on two fixed integer parameters
$s_1,s_2=0,1,...$. 
This new family of models contain interaction four-vertex models ($s_1 
\neq 0, s_2 \neq 0$), interacting five-vertex models ($s_1=0,s_2 \neq 0$, or 
$s_1 \neq 0, s_2 =0$) and the standard six-vertex model ($s_1=s_2=0$).
These vertex models can also be interpreted as if the
vertical arrows entering in the vertex configurations have an effective
hard-core size $s=s_1+s_2+1$. 

For a given value of $s_1$ and $s_2$ the model can be parametrized by four 
parameters ($\gamma, \nu, \mu$ and $h$) (see \rf{r1}). The exact solution of 
the eigenspectra of the row-to-row transfer matrix shows the independence of the 
eigenfunctions on the parameters $\nu$ and $\mu$, implying the commutativity of 
an infinite family of transfer matrices (see \rf{r3} and \rf{r4}).  The expansion of 
the transfer matrix in terms of the free parameter $\mu$ give an infinite set of 
commuting conserved charges. A member of this set  is a XXZ quantum chain 
with hard-core interactions that exclude two up spins ($\sigma^z$-basis) at 
distances smaller than $s_1+s_2+1$ \cite{alcbar1,maquire}. This quantum chain also describes 
the asymmetric diffusion of particles with hard-core size $s_1+s_2+1$, in units 
of lattice spacing \cite{sazamoto,alcbar5,alcbar6}.  
The  integrability of this quantum chain is 
already known \cite{alcbar1,maquire}. The present paper give an explanation for this 
integrability, namely, the existence of an infinite set of commuting charges 
generated by the row-to-row transfer matrix of the introduced 
interacting vertex models. 

The phase diagram of the introduced interacting vertex models, 
as discussed in Sec. $4$, can be obtained from the known results
 of the six-vertex model. 
The critical phases of the models are 
governed by a Coulomb gas type of conformal field theories with central charge 
$c=1$. Since the spectral parameter equations are the same as those of the 
extended XXZ quantum chain \cite{maquire} 
the compactification ratio, that fixes the critical exponents, depend on the 
density of vertical arrows $n$ and on the parameters $s_1$ and $s_2$.

The exact solution we obtained for the interacting vertex models introduced 
in this paper were done by using the matrix product ansatz introduced in 
\cite{alclazo1}-\cite{alclazo4}. The most elegant solution of the six-vertex model is obtained 
by using the $R$-matrix approach in the quantum inverse scattering method. 
Can we solve the introduced interacting vertex models by generalizing the 
$R$ matrix of the six-vertex model? Due to the appearance of nonlocal 
operators in the row-to-row transfer matrix we were not able to find such 
generalization. That is an interesting question to be answered in the 
future. 


Similar to the hard-core generalization of the XXZ quantum chain, has 
been also generalized several known  integrable quantum chains 
through the introduction of suitable hard-core exclusion interactions. 
As examples we have hard-core generalizations of the spin-1 
Fateev-Zamolodchikov and Izergin-Korepin models \cite{alcbar3}, 
$S_qU(N)$ Sutherland and Perk-Schultz models \cite{alcbar6,alcbar7} and 
the Bariev model \cite{schlot1,schlot2}. 
  These quantum 
chains with no hard-core interactions, as the standard XXZ chain, are 
known to be generated by the row-to-row transfer matrix of known vertex 
models. We believe that the same ideas  presented in this paper for the 
generalization of the six-vertex model can also be extended to these 
 vertex models. We expect that the vertex models generating the 
quantum chains with only next neighbour interactions are special cases of
 a larger family of integrable 
interacting vertex models.

\section*{Acknowledgments} 
 This work has been partly supported by FAPESP, CNPq and CAPES (Brazilian agencies).

\end{document}